\documentclass[aps,prd,twocolumn,showpacs,nofootinbib]{revtex4-1}

\usepackage{graphicx}
\usepackage{amsmath}
\usepackage{amssymb}
\usepackage{bm}
\usepackage{bbm}
\usepackage{color}
\usepackage{slashed}
\usepackage[colorlinks=true,linkcolor=blue,citecolor=blue, urlcolor=blue]{hyperref}

\begin{document}
\hyphenpenalty=5000
\tolerance=1000

\title{Production of the $B_c$ meson at the CEPC}

\author{Ze-Yang Zhang}
\email{zhangzeyang@cqu.edu.cn}
\author{Xu-Chang Zheng}
\email{zhengxc@cqu.edu.cn}
\author{Xing-Gang Wu}
\email{wuxg@cqu.edu.cn}

\affiliation{Department of Physics, Chongqing University, Chongqing 401331, P.R. China\\
Chongqing Key Laboratory for Strongly Coupled Physics, Chongqing 401331, P.R. China}

\begin{abstract}

In this paper, we make a detailed study on the production of the $B_c$, $B_c^*$, $B_c(2\,^1S_0)$, and $B^*_c(2\,^3S_1)$ mesons via the three planned running modes ($Z$, $W$, and $H$) at the future Circular Electron-Positron Collider (CEPC). The fragmentation-function and full nonrelativistic QCD approaches are adopted to calculate the production cross sections. Considering the excited states shall decay to the ground $B_c$ state (e.g. $1^1S_0$-state) with almost $100\%$ probability, our numerical results show that up to next-to-leading order QCD corrections, there are about $1.4\times 10^8$ $B_c$ events to be accumulated via the $Z$ mode ($\sqrt{s}=m_{_Z}$) of the CEPC, but only about $1.6\times 10^4$ and $1.1\times 10^4$ $B_c$ events to be accumulated via the $W$ ($\sqrt{s}=160\,{\rm GeV}$) and $H$ ($\sqrt{s}=240\,{\rm GeV}$) modes, respectively. Since the $Z$ mode is the best mode among the three planned modes of the CEPC for studying the production of the $B_c$, $B_c^*$, $B_c(2\,^1S_0)$, and $B^*_c(2\,^3S_1)$ mesons and the differential distributions of these mesons may be measured precisely at this mode, we further present the differential cross sections $d\sigma/(dz\,d{\rm cos}\theta)$ and $d\sigma/dz$ via the $Z$ mode of the CEPC.

\end{abstract}

\maketitle

\section{Introduction}

The $B_c$ meson, which usually stands for the ground $1^1S_0$-state of $(c\bar{b})$-quarkonium and carries two different heavy flavors, provides a unique bound-state system for testing the Standard Model (SM) and has attracted lots of interests. The study on the production of the $B_c$ meson is important for the researches of the $B_c$ meson, since it tells us how many $B_c$ events can be produced in a collider platform. On the other hand, the theoretical models for the $B_c$ production can be tested after fixing its production rate by measuring its cascade decays. Similar to the heavy quarkonia such as $\eta_c$, $J/\psi$, $\eta_b$, $\Upsilon$, and etc., the $B_c$ meson production cross section can also be factorized through the nonrelativistic QCD (NRQCD) \cite{nrqcd} factorization theory and a lot of information for the production of $B_c$ can be calculated perturbatively. Compared with the quarkonium production where color-octet contribution is usually important \cite{Brambilla:2010cs, Andronic:2015wma, Lansberg:2019adr, Chen:2021tmf}, the production mechanism of the $B_c$ meson is simpler due to the fact that the color-octet contribution is always suppressed compared to the color-singlet contribution in the $B_c$ production processes.

At present, most studies of the $B_c$ production focus on the hadronic production due to the high production rate at a hadron collider, such as Tevatron or LHC \cite{Chang:1992jb, Chang:1994aw, Kolodziej:1995nv, Berezhnoy:1994ba, Chang:1996jt, Berezhnoy:1996ks, Baranov:1997wy, Baranov:1997sg, Cheung:1999ir, Chang:2003cr, Chang:2004bh, Chang:2005bf, Chang:2005wd, Chang:2003cq, Chang:2005hq, Wang:2012ah, Chen:2018obq, Berezhnoy:2019yei}. Indeed, the $B_c$ meson has been only observed at hadron colliders up to now \cite{CDF:1998ihx, CDF:1998axz, CDF:2012ksy, D0:2008bqs, D0:2008thm, LHCb:2012ihf, LHCb:2013xlg, LHCb:2013hwj, LHCb:2014ilr, LHCb:2014mvo, LHCb:2017lpu, CMS:2021vuz}. In addition to hadron colliders, a high luminosity $e^+e^-$ collider can be a potentially good platform for studying the $B_c$ meson. Compared with the hadron collider, there are less backgrounds at the $e^+e^-$ collider, thus it is suitable for studying the $B_c$ meson precisely. Moreover, the production mechanism of the $B_c$ meson at the $e^+e^-$ collider is much simpler than that at the hadron collider. Recent studies on the $B_c$ production at the $e^+e^-$ colliders can be found in Refs.\cite{Yang:2011ps, Yang:2013vba, Chen:2014xka, Berezhnoy:2016etd, Zheng:2015ixa, Zheng:2017xgj, Zheng:2018fqv, Chen:2020dtu}.

The Chinese particle physics community has proposed to construct a high luminosity $e^+e^-$ collider in China which is called the Circular Electron-Positron Collider (CEPC) \cite{CEPCStudyGroup:2018ghi}. The CEPC will operate at three modes: $H(e^+e^- \to ZH)$ at around $\sqrt{s}=240 \,{\rm GeV}$, $Z(e^+e^- \to Z)$ at around $\sqrt{s}=m_{_Z}\simeq 91.2 \, {\rm GeV}$, and $W(e^+e^- \to W^+ W^-)$ at around $\sqrt{s}=160 \,{\rm GeV}$. The integrated luminosities of those modes are planned as $5.6 \,ab^{-1}$, $16 \,ab^{-1}$ and $2.6 \,ab^{-1}$, respectively \cite{An:2018dwb}. As programmed, in addition to studying the Higgs boson properties, the CEPC also provides a good platform for studying the heavy flavor physics. Therefore, as one of the hot topics in heavy flavor physics, the study of the $B_c$ production at the CEPC is very interesting. In this paper, we will devote ourselves to studying the properties of the inclusive $B_c$ production at the CEPC.

Since the $B_c$ meson carries two different heavy flavors, its excited states below the BD threshold will directly or indirectly decay to the ground state via the electro-magnetic or strong interactions with a probability of almost $100\%$. These excited states are important sources of the $B_c$ meson. Moreover, the $2S$ excited states $B_c(2 \,^1S_0)$ and $B^*_c(2 \,^3S_1)$ have been recently observed by the CMS and the LHCb collaborations \cite{CMS:2019uhm, LHCb:2019bem}. Thus, the study on the production of the two excited states is interesting by itself. In this paper, besides the production of the ground state $B_c$ meson, we will also consider the production of the excited states $B^*_c(1 \,^3S_1)$, $B_c(2 \,^1S_0)$, and $B^*_c(2 \,^3S_1)$.

\section{Calculation technology}

According to the NRQCD factorization formalism, the production cross section of the $B_c$ meson in $e^+e^-$ collisions can be written as \footnote{For simplicity, we only present the formulas for the ground-state $B_c$ meson, the formulas for the $B^*_c$, $B_c(2\,^1S_0)$, and $B^*_c(2\,^3S_1)$ states are similar to those of the $B_c$ meson.}
\begin{eqnarray}
d \sigma_{e^+ e^- \to B_c+X}=&&\sum_n d \tilde{\sigma}_{e^+e^- \to (c\bar{b})[n]+X}\langle {\cal O}^{B_c}(n) \rangle.
\label{nrqcd}
\end{eqnarray}
where $d \tilde{\sigma}$ denotes the production cross section for the perturbative state $(c\bar{b})[n]$ with the quantum numbers $n$, which can be calculated using perturbative QCD (pQCD). $\langle {\cal O}^{B_c}(n) \rangle$ denotes the long-distance matrix element (LDME) which is proportional to the transition probability for a $(c\bar{b})[n]$ pair into the corresponding $B_c$ meson state. In the lowest nonrelativistic approximation, only the color-singlet LDME need to be considered, which can be related to the wave function at the origin [i.e., $\langle {\cal O} ^{B_c}(^1S_0^{[1]})\rangle \approx N_c \vert R_S(0)\vert^2/(2\pi)=\langle {\cal O} ^{c\bar{b}[^1S_0^{[1]}]}(^1S_0^{[1]})\rangle \vert R_S(0)\vert^2/(4\pi)$] and can be calculated through phenomenological potential models. In addition to the factorization formula (\ref{nrqcd}), when the electron-position collision energy $\sqrt{s} \gg m_{B_c}$, the production cross section can also be factorized as
\begin{eqnarray}
\frac{d^2 \sigma_{e^+ e^- \to B_c+X}}{dz\,d{\rm cos}\theta}=&&\sum_i \int \frac{dy}{y}\frac{d^2 \hat{\sigma}_{e^+e^- \to i+X}}{dy\,d{\rm cos}\theta}(y,\mu_F)\nonumber\\ &&\times D_{i\to B_c}(z/y,\mu_F),
\label{pqcd}
\end{eqnarray}
where $d^2 \hat{\sigma}_{e^+e^- \to i+X}/(dy\,d{\rm cos}\theta)$ denotes the partonic cross section, $D_{i\to B_c}(z/y,\mu_F)$ denotes the fragmentation function for a parton $i$ into the $B_c$ meson, $z=2 p_{B_c}\cdot q/q^2$ (here, $q$ is the sum of initial-state $e^+$ and $e^-$ momenta) is the momentum fraction carried by the $B_c$ meson, $\theta$ is the angle between the momenta of the $B_c$ meson and the initial-state $e^-$, and $\mu_F$ is the factorization scale. This factorization formalism is known as the fragmentation-function approach. Unlike the fragmentation functions for the light hadrons which are nonperturbative in nature, the fragmentation function for the $B_c$ meson contains much information which is perturbatively calculable, e.g., based on the NRQCD factorization,
\begin{eqnarray}
D_{i\to B_c}(z,\mu_F)=\sum_n d_{i\to (c\bar{b})[n]}(z,\mu_F) \langle {\cal O}^{B_c}(n) \rangle,
\label{frag-nrqcd}
\end{eqnarray}
where $d_{i\to (c\bar{b})[n]}(z,\mu_F)$ is the calculable short-distance coefficient. The leading-order (LO) fragmentation functions for the $B_c$ meson and its excited states have been calculated in the 1990s \cite{Chang:1992bb, Braaten:1993jn, Ma:1994zt, Chen:1993ii, Yuan:1994hn, Cheung:1995ir}. The next-to-leading-order (NLO) fragmentation functions $D_{\bar{b} \to B_c}(z,\mu_F)$ and $D_{c \to B_c}(z,\mu_F)$ can be found in Ref.\cite{Zheng:2019gnb}.

One advantage of the fragmentation-function approach is that the large logarithms of $s/m_Q^2$ ($m_Q$ is the heavy quark mass) appearing in the perturbative expansion of the cross section can be resummed through solving the evolution equations of the fragmentation functions. For example, the differential cross section up to NLO and next-to-leading logarithmic (NLL) accuracy can be calculated through
\begin{eqnarray}
\frac{d^2 \sigma^{\rm Frag,NLO+NLL}_{e^+ e^- \to B_c+X}}{dz\,d{\rm cos}\theta}=&&\sum_{i=c,\bar{b}} \int \frac{dy}{y}\frac{d^2 \hat{\sigma}^{\rm NLO}_{e^+e^- \to i+X}}{dy\,d{\rm cos}\theta}(y,\mu_F)\nonumber\\
&&\times D^{\rm NLO+NLL}_{i\to B_c}(z/y,\mu_F),
\label{Frag-NLL}
\end{eqnarray}
where the expression of the partonic cross section up to NLO accuracy can be found in Ref.\cite{Nason:1993xx}. To avoid the large logarithms appearing in the partonic cross section, we shall implicitly set the factorization scale and renormalization scale to be the same, i.e., $\mu_F=\mu_R=\sqrt{s}$.  Then the large logarithms of $s/m_Q^2$ appear in the fragmentation functions. We derive the fragmentation functions $D^{\rm NLO+NLL}_{\bar{b} \to B_c}(z,\mu_F=\sqrt{s})$ and $D^{\rm NLO+NLL}_{c \to B_c}(z,\mu_F=\sqrt{s})$ by solving the DGLAP equation \cite{Dokshitzer:1977sg, Gribov:1972ri, Altarelli:1977zs}, and the NLO fragmentation functions $D^{\rm NLO}_{\bar{b}\to B_c}(z,\mu_{F0}=m_b+2m_c)$ and $D^{\rm NLO}_{c\to B_c}(z,\mu_{F0}=2m_b+m_c)$ with $\mu_R=\mu_{F0}$ obtained in Ref.\cite{Zheng:2019gnb} are used as the boundary condition \footnote{To avoid the large logarithms appearing in the initial fragmentation functions, the factorization and renormalization scales of the initial fragmentation functions are set to be the minimal invariant mass of the fragmentation quark, i.e., $\mu_{F0}=\mu_R=m_b+2m_c$ for the $\bar{b}$-quark fragmentation channel and $\mu_{F0}=\mu_R=2m_b+m_c$ for the $c$-quark fragmentation channel.}. For solving the DGLAP equation, we adopt the Mellin transformation method, and the NLO splitting function $P_{QQ}(y)$ \cite{Curci:1980uw,Furmanski:1980cm,Floratos:1978ny,Gonzalez-Arroyo:1979qht,Floratos:1981hs} is used as the evolution kernel. The related formulas for the Mellin transformation of the NLO splitting function can be found in Ref.\cite{Mele:1990cw}, and the inverse Mellin transformation can be carried out numerically \cite{Graudenz:1995sk}. Through the DGLAP equation, the large logarithms of $s/m_Q^2$ in the fragmentation functions are resummed to NLL accuracy.

As a comparison, we also calculate the differential cross section at the NLO level alone under the fragmentation-function approach, i.e., without doing the resummation of the large logarithms of $s/m_Q^2$,
\begin{eqnarray}
\frac{d^2 \sigma^{\rm Frag,NLO}_{e^+ e^- \to B_c+X}}{dz\,d{\rm cos}\theta}=&&\sum_{i=c,\bar{b}} \int \frac{dy}{y}\frac{d^2 \hat{\sigma}^{\rm NLO}_{e^+e^- \to i+X}}{dy\,d{\rm cos}\theta}(y,\mu_{F})\nonumber\\
&&\times D^{\rm NLO}_{i\to B_c}(z/y,\mu_{F}),
\label{Frag-NLO}
\end{eqnarray}
where the factorization and renormalization scales of the partonic cross section and the fragmentation functions are set as the minimal invariant mass of the fragmentation quark, i.e., $\mu_F=\mu_R=m_b+2m_c$ for the $\bar{b}$-quark fragmentation channel and $\mu_F=\mu_R=2m_b+m_c$ for the $c$-quark fragmentation channel.

\section{Numerical results and discussions}

To do the numerical calculation, the necessary input parameters are taken as follows:
\begin{eqnarray}
&& m_c=1.5\,{\rm GeV},\, m_b=4.9\,{\rm GeV},\, m_{_Z}=91.1876\,{\rm GeV},\nonumber \\
&& {\rm sin}^2\theta_w=0.231,\;\alpha=1/128,\; \Gamma_{_Z}=2.4952\,{\rm GeV},\nonumber \\
&& |R_{1S}(0)|^2=1.642\,{\rm GeV}^3,\, |R_{2S}(0)|^2=0.983\,{\rm GeV}^3.
\end{eqnarray}
Here $\alpha=\alpha(m_{_Z})$ is the electromagnetic coupling constant at the scale $m_{_Z}$, $|R_{1S}(0)|$ and $|R_{2S}(0)|$ are radial wave functions at the origin for the color-singlet $1S$-level and $2S$-level $(c\bar{b})$-quarkonium states, which are taken from the calculation based on the Buchmuller-Tye potential model~\cite{Eichten:1994gt}. For the strong coupling constant, we adopt the two-loop formula:
\begin{equation}
\alpha_s(\mu)=\frac{4\pi}{\beta_0~{\rm ln}(\mu^2/\Lambda^2_{\rm QCD})}\left[ 1-\frac{\beta_1~{\rm ln}~{\rm ln}(\mu^2/\Lambda^2_{\rm QCD})}{\beta_0^2~{\rm ln}(\mu^2/\Lambda^2_{\rm QCD})}\right],
\end{equation}
where $\beta_0=11-\frac{2}{3} n_f$ and $\beta_1=102-\frac{38}{3} n_f$ ($n_f$ being the active flavor numbers) are the one-loop and two-loop $\{\beta_i\}$-functions. According to $\alpha_s(m_{_Z})=0.1185$ \cite{ParticleDataGroup:2016lqr}, we obtain $\Lambda_{\rm QCD}^{n_f=5}=0.233\, {\rm GeV}$.

\subsection{Basic results for the three production modes at the CEPC under the fragmentation-function approach up to NLO+NLL level}

\begin{table}[htb]
\begin{tabular}{c c c c}
\hline
~States~  & ~$\sqrt{s}=m_{_Z}$~& ~$\sqrt{s}=160\,{\rm GeV}$~ & ~$\sqrt{s}=240\,{\rm GeV}$~   \\
\hline
$B_c$ & 2.58 & $1.71\times 10^{-3}$ &  $5.34\times 10^{-4}$ \\
$B^*_c$  & 3.04 & $2.05\times 10^{-3}$ &  $6.40\times 10^{-4}$ \\
$B_c(2^1S_0)$  & 1.54 & $1.03\times 10^{-3}$ &  $ 3.20\times 10^{-4}$ \\
$B^*_c(2^3S_1)$ & 1.82 & $1.23\times 10^{-3}$ &  $ 3.83\times 10^{-4}$ \\
\hline
\end{tabular}
\caption{The cross sections (in unit: $pb$) for the production of the $B_c$, $B_c^*$, $B_c(2\,^1S_0)$, and $B_c^*(2\,^3S_1)$ states at $\sqrt{s}=m_{_Z}$, $\sqrt{s}=160\,{\rm GeV}$, and $\sqrt{s}=240\,{\rm GeV}$ under the fragmentation-function approach up to NLO+NLL level.}
\label{tb-section}
\end{table}

The cross sections for the production of $B_c$, $B_c^*$, $B_c(2\,^1S_0)$, and $B_c^*(2\,^3S_1)$ states via the three modes of the CEPC under the fragmentation-function approach up to NLO+NLL level are given in TABLE \ref{tb-section}. From the table, we can see that the cross sections for the production of the $B_c$, $B_c^*$, $B_c(2\,^1S_0)$, and $B_c^*(2\,^3S_1)$ states at $\sqrt{s}=m_{_Z}$ are larger than the corresponding cross sections at $\sqrt{s}=160\,{\rm GeV}$ and $\sqrt{s}=240\,{\rm GeV}$ by about three to four orders of magnitude. This is because the cross sections are enhanced by the $Z$-boson resonance effect when $\sqrt{s}=m_{_Z}$. More explicitly, the squared amplitude of the $Z$-boson exchange diagrams is proportional to a factor of $1/[(s-m_{_Z}^2)^2+m_{_Z}^2 \Gamma_Z^2]$.  This factor has a peak at the $Z$ pole, and it decreases rapidly when $\sqrt{s}$ deviates from $m_{_Z}$. It has the value of $1.93\times 10^{-5}\, {\rm GeV}^{-4}$ (as a comparison, the corresponding factor $1/s^2$ of the $\gamma$-exchange diagrams is $1.45\times 10^{-8}\, {\rm GeV}^{-4}$) at the $Z$ pole, which quickly decreases to $3.35\times 10^{-9}\, {\rm GeV}^{-4}$ and $4.12\times 10^{-10}\, {\rm GeV}^{-4}$ when $\sqrt{s}=160\,{\rm GeV}$ and $\sqrt{s}=240\,{\rm GeV}$, respectively.

\begin{table}[htb]
\begin{tabular}{c c c c}
\hline\hline
~States~  & ~$\sqrt{s}=m_{_Z}$~& ~$\sqrt{s}=160\,{\rm GeV}$~ & ~$\sqrt{s}=240\,{\rm GeV}$~   \\
\hline
$B_c$ & $4.13\times10^7$ & $4.45\times10^3$  & $2.99\times10^3$ \\
$B^*_c$  & $4.86\times10^7$ & $5.33\times10^3$  &  $3.58\times10^3$ \\
$B_c(2^1S_0)$  & $2.46\times10^7$ & $2.68\times10^3$ & $1.79\times10^3$  \\
$B^*_c(2^3S_1)$ & $2.91\times10^7$ & $3.20\times10^3$ &  $2.14\times10^3$ \\
\hline\hline
\end{tabular}
\caption{The expected numbers of the $B_c$, $B_c^*$, $B_c(2\,^1S_0)$, and $B_c^*(2\,^3S_1)$ events to be generated via the three planned modes of the CEPC.}
\label{tb-events}
\end{table}

With the planned luminosities of the three running modes at the CEPC, the numbers of the events of these states to be generated via the three modes of the CEPC are estimated and presented in Table \ref{tb-events}. We can see that there are ${\cal O}(10^7)$ [${\cal O}(10^3)$, ${\cal O}(10^3)$] $B_c$, $B_c^*$, $B_c(2\,^1S_0)$, and $B_c^*(2\,^3S_1)$ events to be produced at the $Z$ ($W$, $H$) mode of the CEPC. Therefore, the CEPC (especially the planned $Z$ mode) will provide a new opportunity for studying the properties of the $B_c$ meson.

\subsection{More studies on the $Z$ mode of the CEPC}

Since the $Z$-boson resonance effect shall raise the production rates by several orders of magnitude at the $Z$ pole, the total production cross sections as well as the differential distributions of the $B_c$, $B_c^*$, $B_c(2\,^1S_0)$, and $B_c^*(2\,^3S_1)$ states may be measured precisely via the $Z$ mode of the CEPC. Therefore, it is important to give a more detailed study on the production of those states at the $Z$ pole.

\begin{figure}[htbp]
\includegraphics[width=0.5\textwidth]{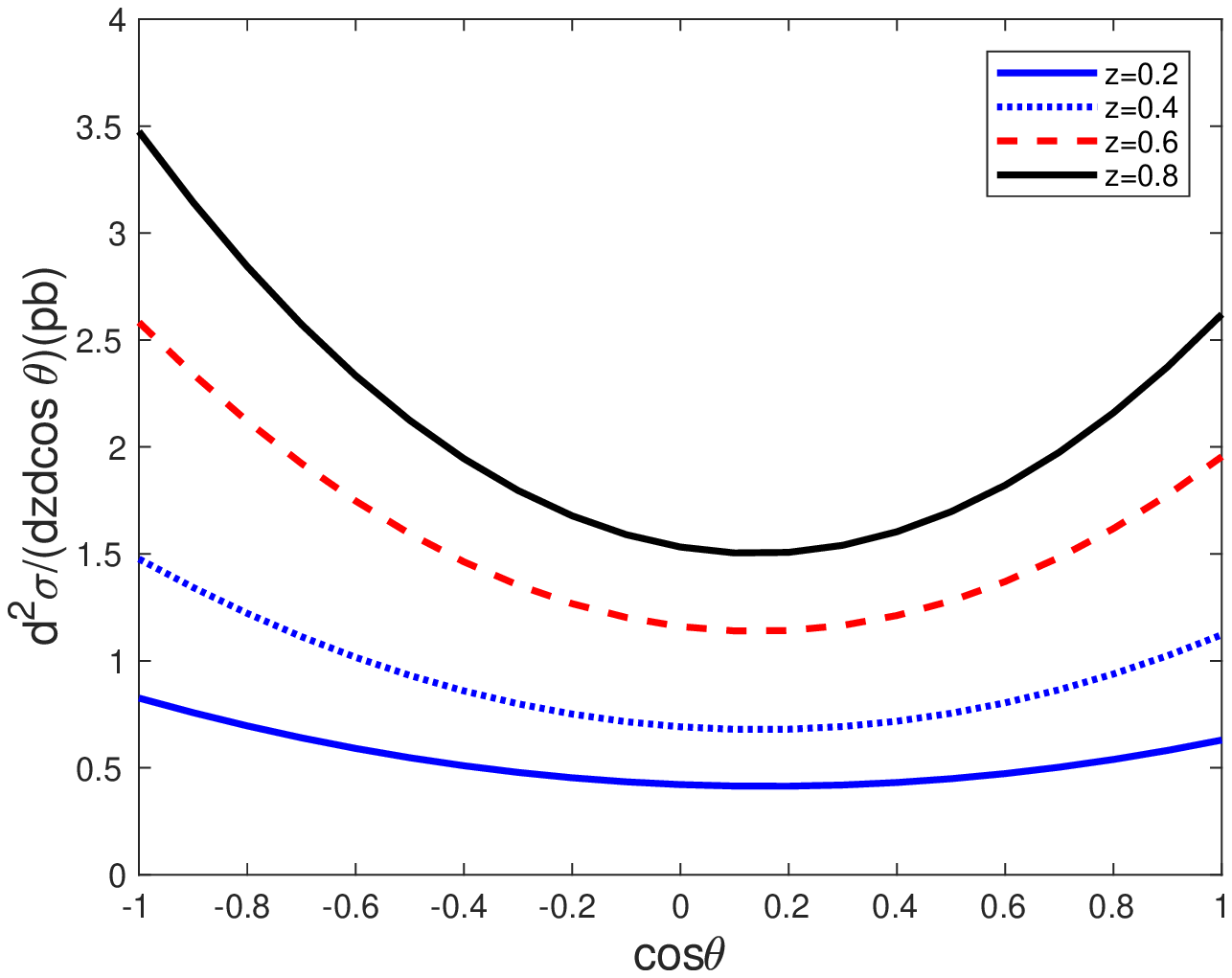}
\caption{The differential cross section $d^2 \sigma/(dz\,d{\rm cos}\theta)$ for the production of the $B_c(1\,^{1}S_0)$ meson at the $Z$ pole under the fragmentation-function approach up to NLO+NLL level, where $\theta$ is the angle between the momenta of the final $B_c$ meson and the initial electron.} \label{cos1s0}
\end{figure}

\begin{figure}[htbp]
\includegraphics[width=0.5\textwidth]{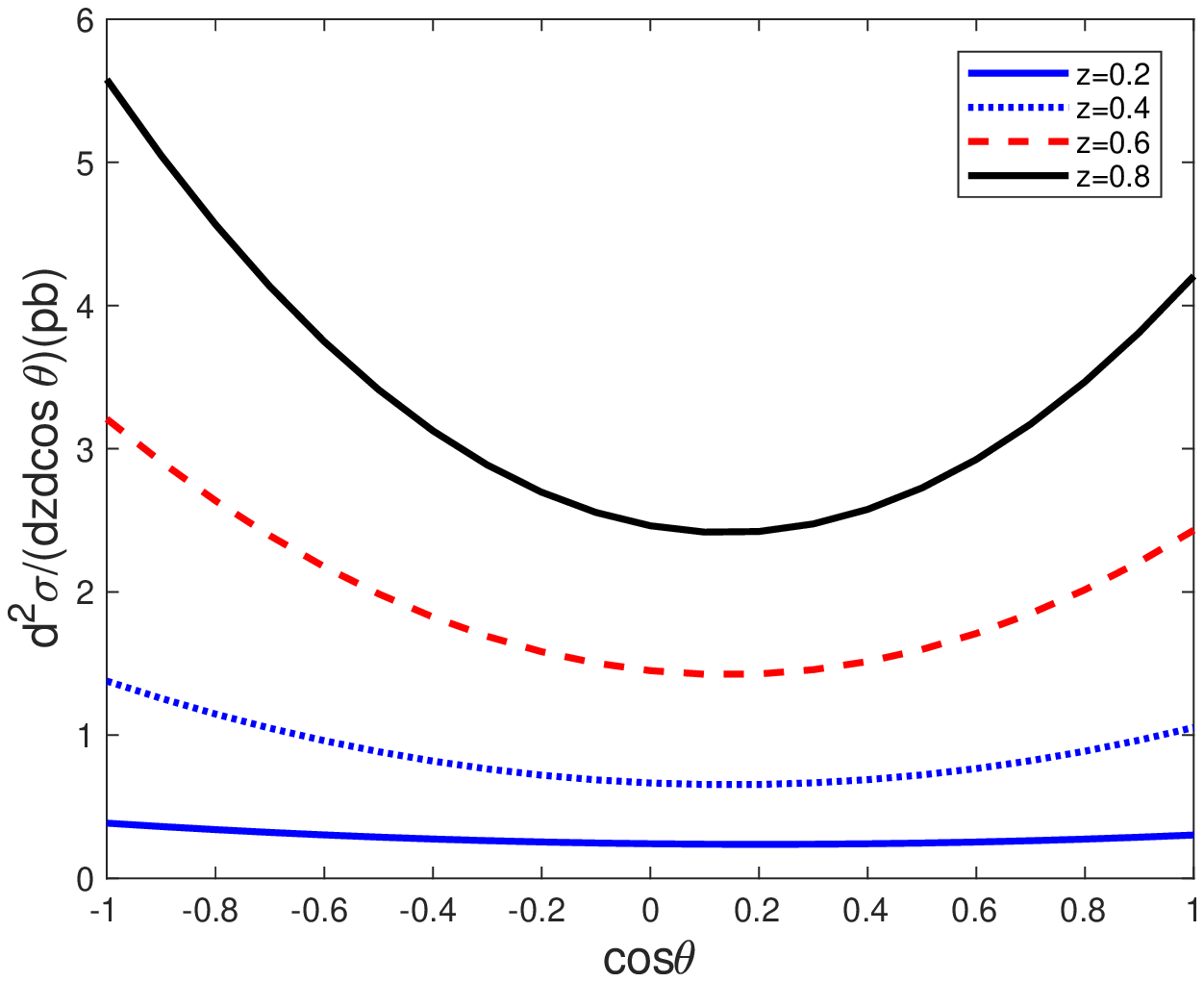}
\caption{The differential cross section $d^2 \sigma/(dz\,d{\rm cos}\theta)$ for the production of the $B_c^*(1\,^{3}S_1)$ meson at the $Z$ pole under the fragmentation-function approach up to NLO+NLL level, where $\theta$ is the angle between the momenta of the final $B_c^*$ meson and the initial electron.} \label{cos3s1}
\end{figure}

\begin{figure}[htbp]
\includegraphics[width=0.5\textwidth]{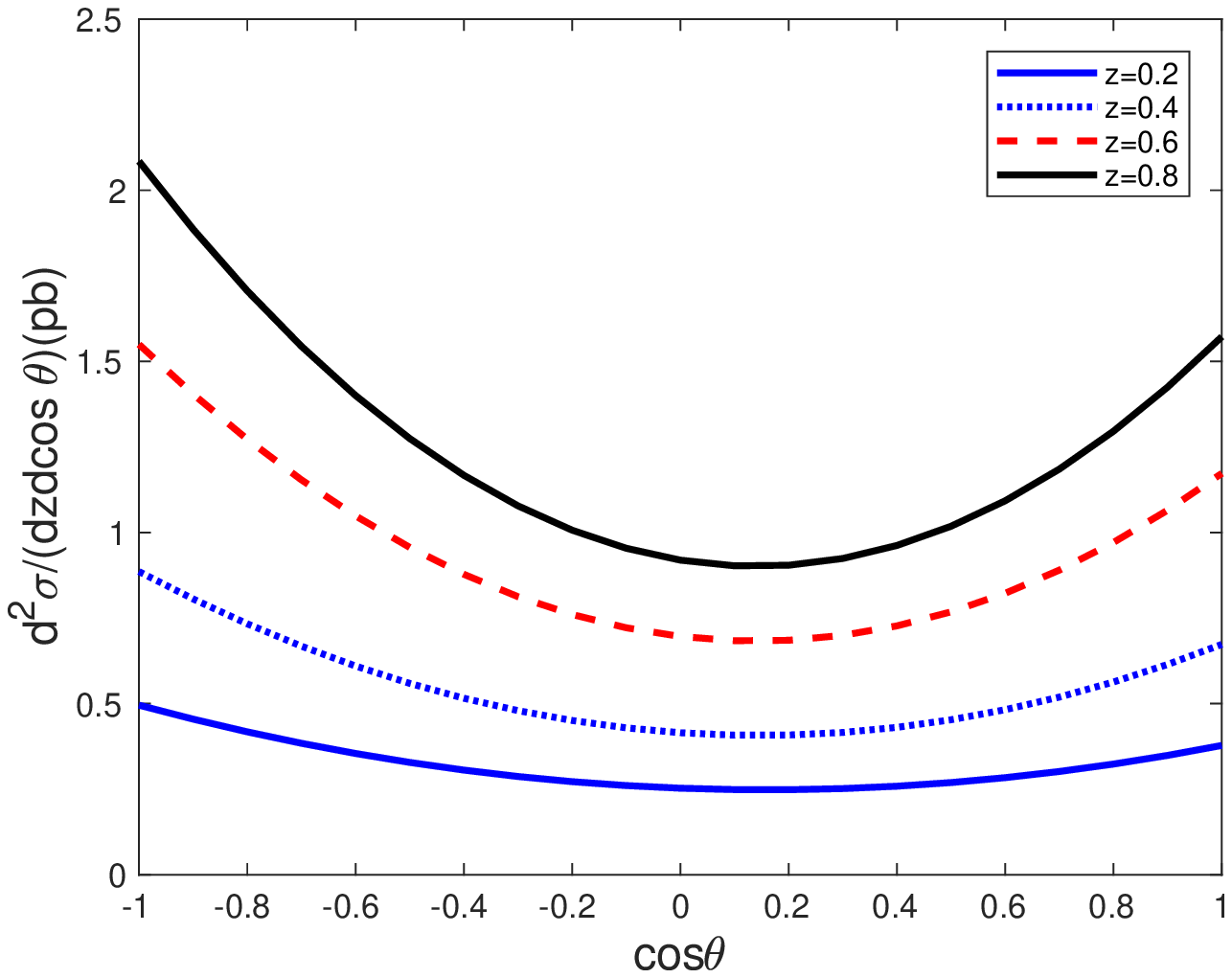}
\caption{The differential cross section $d^2 \sigma/(dz\,d{\rm cos}\theta)$ for the production of the $B_c(2\,^1S_0)$ meson at the $Z$ pole under the fragmentation-function approach up to NLO+NLL level, where $\theta$ is the angle between the momenta of the final $B_c(2\,^1S_0)$ meson and the initial electron.} \label{cos1s02}
\end{figure}

\begin{figure}[htbp]
\includegraphics[width=0.5\textwidth]{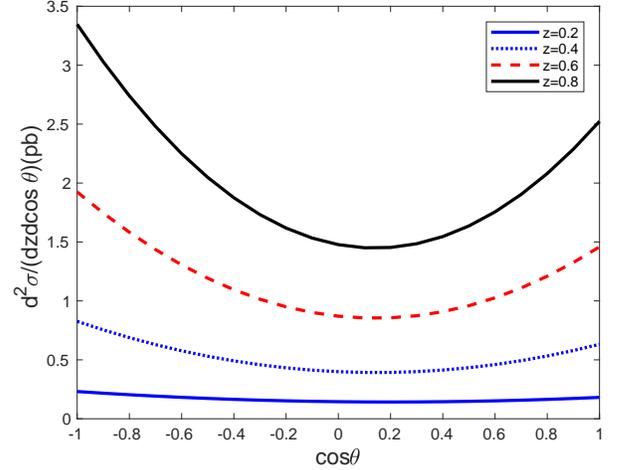}
\caption{The differential cross section $d^2 \sigma/(dz\,d{\rm cos}\theta)$ for the production of the $B_c^*(2\,^3S_1)$ meson at the $Z$ pole under the fragmentation-function approach up to NLO+NLL level, where $\theta$ is the angle between the momenta of the final $B_c^*(2\,^3S_1)$ meson and the initial electron.} \label{cos3s12}
\end{figure}

We present differential cross sections $d^2 \sigma/(dz\,d{\rm cos}\theta)$ by using the fragmentation-function approach up to NLO+NLL level for the production of $B_c$, $B_c^*$, $B_c(2\,^1S_0)$, and $B_c^*(2\,^3S_1)$ states at the $Z$ pole of CEPC in Figs.(\ref{cos1s0}, \ref{cos3s1}, \ref{cos1s02}, \ref{cos3s12}). Results for several typical momentum fractions (e.g. $z=0.2$, 0.4, 0.6 and 0.8) are presented. From those figures, we can see that the differential distributions $d^2 \sigma/(dz\,d{\rm cos}\theta)$ are asymmetric. Those forward-backward asymmetries arise from the parity violation due to the exchange of $Z$ boson.

\begin{figure}[htbp]
\includegraphics[width=0.5\textwidth]{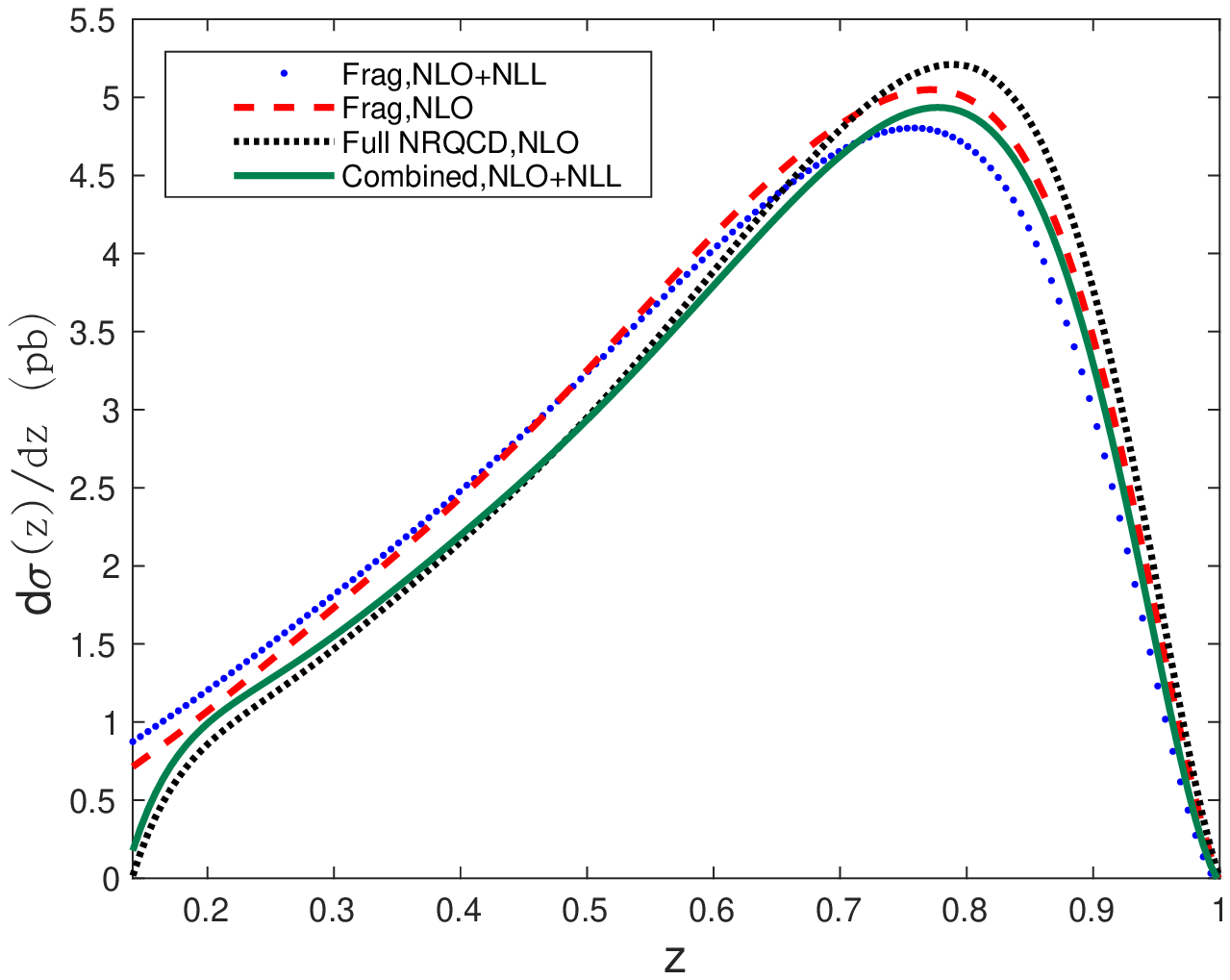}
\caption{The differential cross sections $d\sigma/dz$ for the production of the $B_c$ meson at the $Z$ pole under different approaches.} \label{sigmaz1s01s}
\end{figure}

\begin{figure}[htbp]
\includegraphics[width=0.5\textwidth]{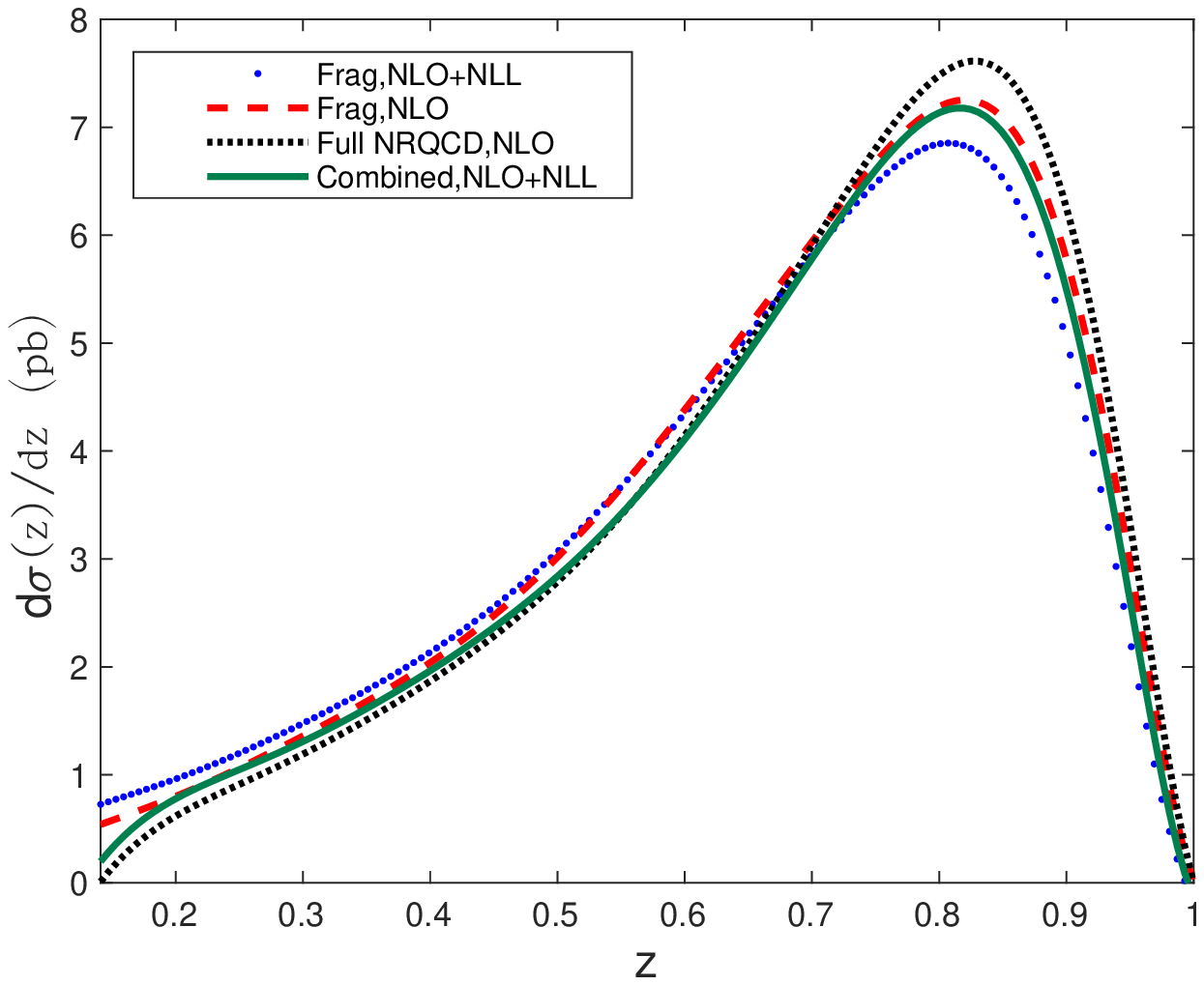}
\caption{The differential cross sections $d\sigma/dz$ for the production of the $B_c^*$ meson at the $Z$ pole under different approaches.} \label{sigmaz3s11s}
\end{figure}

\begin{figure}[htbp]
\includegraphics[width=0.5\textwidth]{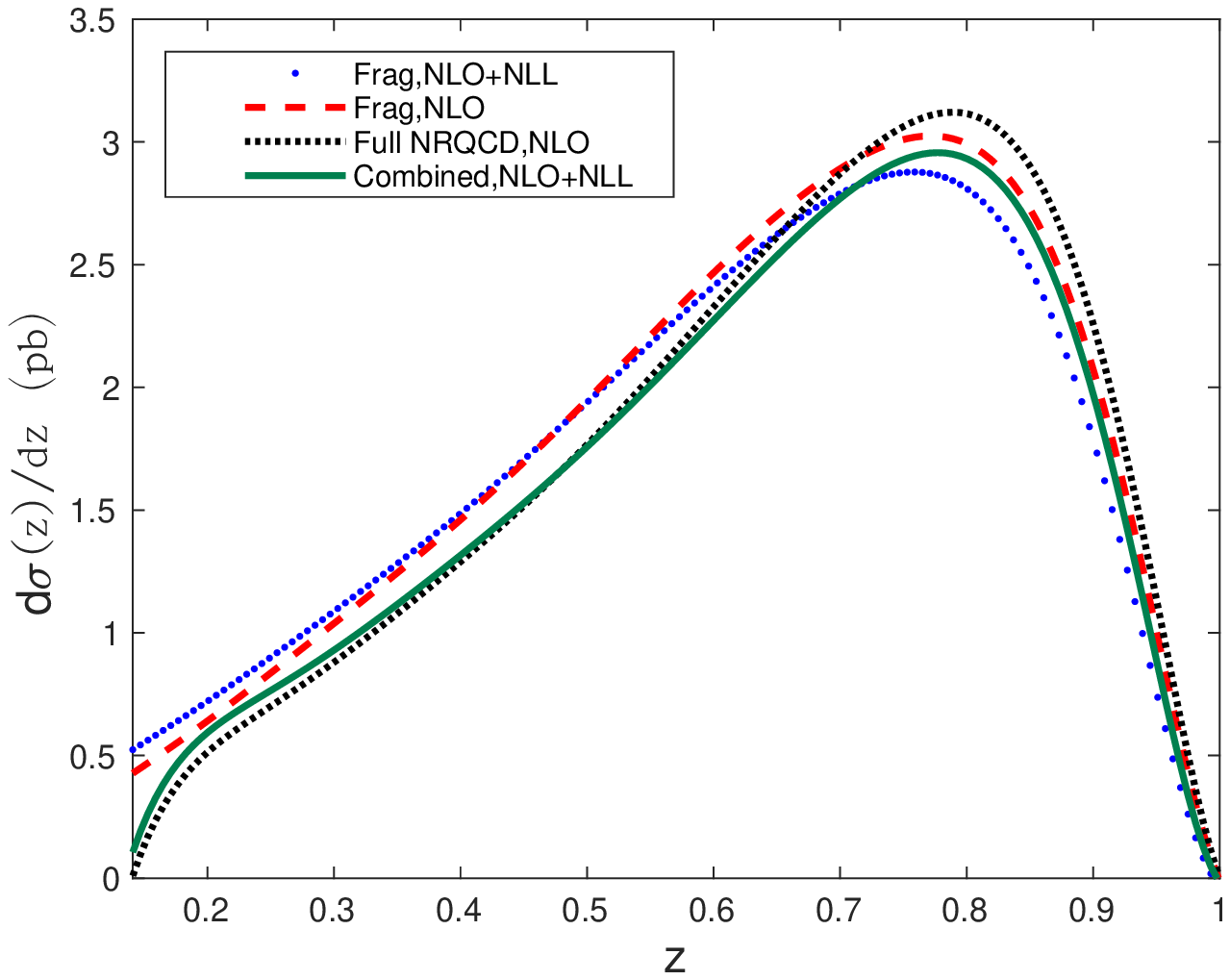}
\caption{The differential cross sections $d\sigma/dz$ for the production of the $B_c(2\,^1S_0)$ meson at the $Z$ pole under different approaches.} \label{sigmaz1s02s}
\end{figure}

\begin{figure}[htbp]
\includegraphics[width=0.5\textwidth]{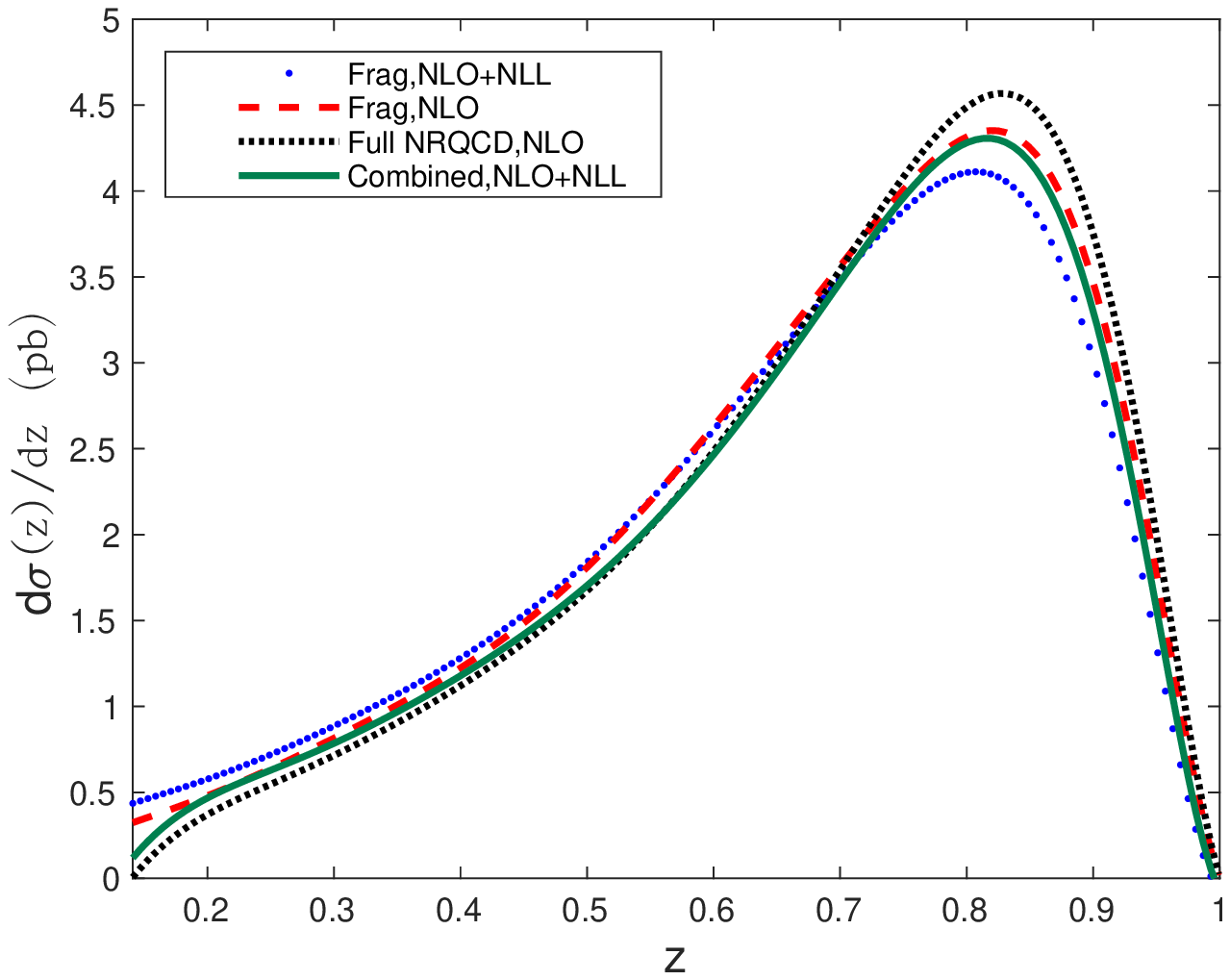}
\caption{The differential cross sections $d\sigma/dz$ for the production of the $B_c^*(2\,^3S_1)$ meson at the $Z$ pole under different approaches.} \label{sigmaz3s12s}
\end{figure}

By further integrating the differential cross sections $d^2 \sigma/(dz\,d{\rm cos}\theta)$ over ${\rm cos}\theta$, we obtain the differential cross sections $d\sigma/dz$, which are presented in Figs.(\ref{sigmaz1s01s}, \ref{sigmaz3s11s}, \ref{sigmaz1s02s}, \ref{sigmaz3s12s}). As a comparison of the NLO+NLL fragmentation-function result (labelled as ``Frag, NLO+NLL"), we also present the results for the full NLO NRQCD approach (labelled as ``Full NRQCD, NLO"), the fragmentation-function approach up to NLO level but without doing the resummation of the large logarithms of $m_Q^2/s$ (labelled as ``Frag, NLO"), and the combined results of the fragmentation-function approach and the full NRQCD approach (labelled as ``Combined, NLO+NLL").

In the fragmentation-function approach, because of the approximation $\sqrt{s} \gg m_{Q}$, some terms which are suppressed by various powers of $m_Q^2/s$ have been neglected. Those neglected terms can be recovered with the help of Eqs.(\ref{nrqcd}, \ref{Frag-NLO}), and we then obtain the combined result of the fragmentation-function approach and the full NRQCD approach up to NLO level, i.e.
\begin{eqnarray}
&&\frac{d^2 \sigma^{\rm Combined,NLO+NLL}_{e^+ e^- \to B_c+X}}{dz\,d{\rm cos}\theta}\nonumber \\
&=& \frac{d^2 \sigma^{\rm Frag, NLO+NLL}_{e^+ e^- \to B_c+X}}{dz\,d{\rm cos}\theta} \nonumber\\
&& + \left(\frac{d^2 \sigma^{\rm Full \,NRQCD, NLO}_{e^+ e^- \to B_c+X}}{dz\,d{\rm cos}\theta} - \frac{d^2 \sigma^{\rm Frag, NLO}_{e^+ e^- \to B_c+X}}{dz\,d{\rm cos}\theta} \right),
\label{FO-Frag-Comb}
\end{eqnarray}
in which the terms given in the parentheses count the contribution neglected in the fragmentation-function formula (\ref{Frag-NLL}). As a subtle point, to be self-consistent with $d^2 \sigma^{\rm Frag, NLO}_{e^+ e^- \to B_c+X}/(dz\,d{\rm cos}\theta)$, the renormalization scale of the differential cross section $d^2 \sigma^{\rm Full \,NRQCD, NLO}_{e^+ e^- \to B_c+X}/(dz\,d{\rm cos}\theta)$, which is calculated through the formula (\ref{nrqcd}) up to NLO accuracy, is taken as the same as that of the dominant $\bar{b}$-fragmentation channel, i.e., $\mu_R=m_b+2m_c$. The combined result (\ref{FO-Frag-Comb}) is more accurate than the results under the fragmentation-function approach or the full NRQCD approach.
From those figures, one can see that the curves from the ``Frag, NLO" are close to the corresponding curves from the ``Full NRQCD, NLO". Compared with the ``Full NRQCD, NLO" results, the ``Frag, NLO" results neglect all the non-leading power contributions. Thus, one can conclude that those contributions are small for the production of the $B_c$, $B_c^*$, $B_c(2\,^1S_0)$, and $B_c^*(2\,^3S_1)$ at the $Z$ pole, and the fragmentation-function approach can give a good approximation to the full NRQCD calculations. The effect of resumming the leading and next-to-leading logarithms of $s/(m_b+2m_c)^2$ or $s/(2m_b+m_c)^2$ can be seen through comparing the ``Frag, NLO+NLL" curves with the ``Frag, NLO" curves. One may observe that after resumming the leading and next-to-leading logarithms of $s/(m_b+2m_c)^2$ or $s/(2m_b+m_c)^2$, the $z$-distributions become softer. However, the effects of the DGLAP evolution for the NLO calculations are not as large as the LO case~\cite{Zheng:2015ixa}. This is because the ``Frag, NLO" calculations contain the first term of the leading logarithmic series, and the effect of the DGLAP evoluntion for the NLO calculation is to resum the leading and next-to-leading logarithms from the corrections beyond the NLO QCD calculations. The ``Combined, NLO+NLL" results not only include the power suppressed terms which have been neglected in the fragmentation-function approach but also resum the leading and next-to-leading logarithms to all orders. Therefore, the ``Combined, NLO+NLL" results are the most accurate results among those predictions shown in Figs.(\ref{sigmaz1s01s}, \ref{sigmaz3s11s}, \ref{sigmaz1s02s}, \ref{sigmaz3s12s}).

Total production cross sections at the $Z$ pole under the ``Combined, NLO+NLL" approach can be obtained through integrating the differential cross sections over $z$. And we obtain $\sigma^{\rm Combined, NLO+NLL}_{e^+ e^- \to B_c+X}=2.51\,{\rm pb}$, $\sigma^{\rm Combined, NLO+NLL}_{e^+ e^- \to B_c^*+X}=3.02\,{\rm pb}$, $\sigma^{\rm Combined, NLO+NLL}_{e^+ e^- \to B_c(2^1S_0)+X}=1.49\,{\rm pb}$, and $\sigma^{\rm Combined, NLO+NLL}_{e^+ e^- \to B_c^*(2^3S_1)+X}=1.80\,{\rm pb}$ at the $Z$ pole.

\section{Summary}

As a summary, we have studied the production of the $B_c$, $B_c^*$, $B_c(2\,^1S_0)$, and $B_c^*(2\,^3S_1)$ mesons at the CEPC. The cross sections for these states via the three planned modes ($Z$, $W$, and $H$) of the CEPC are presented, and the numbers of the events are estimated. If assuming the $B_c^*$, $B_c(2\,^1S_0)$, and $B_c^*(2\,^3S_1)$ states decay to the ground-state $B_c$ with a probability of 100\% via electromagnetic or hadronic interactions, then there are about $1.4 \times 10^8$ ($1.6 \times 10^4$, $1.1 \times 10^4$) $B_c$ events to be produced via the $Z$ ($W$, $H$) mode of the CEPC. Thus we may have a good opportunity to study the properties of the $B_c$ meson via the $Z$ mode of the CEPC. The differential angle and energy distributions for these states via the $Z$ mode are further analyzed. There are relatively fewer $B_c$ events to be produced via the other two ($W$ and $H$) modes. If one wants to study the properties of the $B_c$ meson via the two modes, it is better to improve the luminosities of the two modes properly.

\hspace{2cm}

\noindent {\bf Acknowledgments:} This work was supported in part by the Natural Science Foundation of China under Grants No. 11625520, No. 12005028, No. 12175025, and No. 12147102, by the China Postdoctoral Science Foundation under Grant No. 2021M693743, by the Fundamental Research Funds for the Central Universities under Grant No. 2020CQJQY-Z003, and by the Chongqing Graduate Research and Innovation Foundation under Grant No. ydstd1912.


\begin{thebibliography}{1}

\bibitem{nrqcd}
G.T. Bodwin, E. Braaten and G.P. Lepage,
Rigorous QCD analysis of inclusive annihilation and production of heavy quarkonium,
Phys. Rev. D {\bf 51}, 1125 (1995).

\bibitem{Brambilla:2010cs}
N.~Brambilla, S.~Eidelman, B.~K.~Heltsley, R.~Vogt, G.~T.~Bodwin, E.~Eichten, A.~D.~Frawley, A.~B.~Meyer, R.~E.~Mitchell and V.~Papadimitriou, \textit{et al.}
Heavy Quarkonium: Progress, Puzzles, and Opportunities,
Eur. Phys. J. C \textbf{71}, 1534 (2011).

\bibitem{Andronic:2015wma}
A.~Andronic, F.~Arleo, R.~Arnaldi, A.~Beraudo, E.~Bruna, D.~Caffarri, Z.~C.~del Valle, J.~G.~Contreras, T.~Dahms and A.~Dainese, \textit{et al.}
Heavy-flavour and quarkonium production in the LHC era: from proton\textendash{}proton to heavy-ion collisions,
Eur. Phys. J. C \textbf{76}, 107 (2016).

\bibitem{Lansberg:2019adr}
J.~P.~Lansberg,
New Observables in Inclusive Production of Quarkonia,
Phys. Rept. \textbf{889}, 1 (2020).

\bibitem{Chen:2021tmf}
A.~P.~Chen, Y.~Q.~Ma and H.~Zhang,
A short theoretical review of charmonium production,
arXiv:2109.04028 [hep-ph].

\bibitem{Chang:1992jb}
C.~H.~Chang and Y.~Q.~Chen,
The hadronic production of the B(c) meson at Tevatron, CERN LHC and SSC,
Phys. Rev. D \textbf{48}, 4086 (1993).

\bibitem{Chang:1994aw}
C.~H.~Chang, Y.~Q.~Chen, G.~P.~Han and H.~T.~Jiang,
On hadronic production of the B(c) meson,
Phys. Lett. B \textbf{364}, 78 (1995).

\bibitem{Kolodziej:1995nv}
K.~Kolodziej, A.~Leike and R.~Ruckl,
Production of B(c) mesons in hadronic collisions,
Phys. Lett. B \textbf{355}, 337 (1995).

\bibitem{Berezhnoy:1994ba}
A.~V.~Berezhnoy, A.~K.~Likhoded and M.~V.~Shevlyagin,
Hadronic production of B(c) mesons,
Phys. Atom. Nucl. \textbf{58}, 672 (1995).

\bibitem{Chang:1996jt}
C.~H.~Chang, Y.~Q.~Chen and R.~J.~Oakes,
Comparative study of the hadronic production of B(c) mesons,
Phys. Rev. D \textbf{54}, 4344 (1996).

\bibitem{Berezhnoy:1996ks}
A.~V.~Berezhnoy, V.~V.~Kiselev and A.~K.~Likhoded,
Hadronic production of S and P wave states of anti-b c quarkonium,
Z. Phys. A \textbf{356}, 79 (1996).

\bibitem{Baranov:1997wy}
S.~P.~Baranov,
Pair production of B(c)* mesons in p p and gamma gamma collisions,
Phys. Rev. D \textbf{55}, 2756 (1997).

\bibitem{Baranov:1997sg}
S.~P.~Baranov,
Semiperturbative and nonperturbative production of hadrons with two heavy flavors,
Phys. Rev. D \textbf{56}, 3046 (1997).

\bibitem{Cheung:1999ir}
K.~m.~Cheung,
$B_c$ meson production at the Tevatron revisited,
Phys. Lett. B \textbf{472}, 408 (2000).

\bibitem{Chang:2003cr}
C.~H.~Chang and X.~G.~Wu,
Uncertainties in estimating hadronic production of the meson $B_c$ and comparisons between TEVATRON and LHC,
Eur. Phys. J. C \textbf{38}, 267 (2004).

\bibitem{Chang:2004bh}
C.~H.~Chang, J.~X.~Wang and X.~G.~Wu,
Hadronic production of the P-wave excited $B_c$ -states B*(cJ, L=1),
Phys. Rev. D \textbf{70}, 114019 (2004).

\bibitem{Chang:2005bf}
C.~H.~Chang, C.~F.~Qiao, J.~X.~Wang and X.~G.~Wu,
The Color-octet contributions to P-wave $B_c$ meson hadroproduction,
Phys. Rev. D \textbf{71}, 074012 (2005).

\bibitem{Chang:2005wd}
C.~H.~Chang, C.~F.~Qiao, J.~X.~Wang and X.~G.~Wu,
Hadronic production of $B_c (B^*_c)$ meson induced by the heavy quarks inside the collision hadrons,
Phys. Rev. D \textbf{72}, 114009 (2005).

\bibitem{Chang:2003cq}
C.~H.~Chang, C.~Driouichi, P.~Eerola and X.~G.~Wu,
BCVEGPY: An Event generator for hadronic production of the $B_c$ meson,
Comput. Phys. Commun. \textbf{159}, 192 (2004).

\bibitem{Chang:2005hq}
C.~H.~Chang, J.~X.~Wang and X.~G.~Wu,
BCVEGPY2.0: A Upgrade version of the generator BCVEGPY with an addendum about hadroproduction of the P-wave B(c) states,
Comput. Phys. Commun. \textbf{174}, 241 (2006).

\bibitem{Wang:2012ah}
X.~Y.~Wang and X.~G.~Wu,
A Trick to Improve the Efficiency of Generating Unweighted $B_c$ Events from BCVEGPY,
Comput. Phys. Commun. \textbf{183}, 442 (2012).

\bibitem{Chen:2018obq}
G.~Chen, C.~H.~Chang and X.~G.~Wu,
$B_c (B_c^*)$ meson production via the proton-nucleus and the nucleus-nucleus collision modes at the colliders RHIC and LHC,
Phys. Rev. D \textbf{97}, 114022 (2018).

\bibitem{Berezhnoy:2019yei}
A.~V.~Berezhnoy, I.~N.~Belov, A.~K.~Likhoded and A.~V.~Luhinsky,
$B_c$ excitations at LHC experiments,
Mod. Phys. Lett. A \textbf{34}, 1950331 (2019).

\bibitem{CDF:1998ihx}
F.~Abe \textit{et al.} [CDF],
Observation of the $B_c$ meson in $p\bar{p}$ collisions at $\sqrt{s} = 1.8$ TeV,
Phys. Rev. Lett. \textbf{81}, 2432 (1998).

\bibitem{CDF:1998axz}
F.~Abe \textit{et al.} [CDF],
Observation of $B_c$ mesons in $p\bar{p}$ collisions at $\sqrt{s} = 1.8$ TeV,
Phys. Rev. D \textbf{58}, 112004 (1998).

\bibitem{CDF:2012ksy}
T.~Aaltonen \textit{et al.} [CDF],
Measurement of the $B_c^{-}$ meson lifetime in the decay $B_{c}^{-} \rightarrow J/\psi~\pi^{-}$,
Phys. Rev. D \textbf{87}, 011101 (2013).

\bibitem{D0:2008bqs}
V.~M.~Abazov \textit{et al.} [D0],
Observation of the $B_c$ Meson in the Exclusive Decay $B_c \to J/\psi \pi$,
Phys. Rev. Lett. \textbf{101}, 012001 (2008).

\bibitem{D0:2008thm}
V.~M.~Abazov \textit{et al.} [D0],
Measurement of the lifetime of the $B_c^\pm$ meson in the semileptonic decay channel,
Phys. Rev. Lett. \textbf{102}, 092001 (2009).

\bibitem{LHCb:2012ihf}
R.~Aaij \textit{et al.} [LHCb],
Measurements of $B_c^+$ production and mass with the $B_c^+ \to J/\psi \pi^+$ decay,
Phys. Rev. Lett. \textbf{109}, 232001 (2012).

\bibitem{LHCb:2013xlg}
R.~Aaij \textit{et al.} [LHCb],
Observation of the Decay $B^+_c \to B^0_s \pi^+$,
Phys. Rev. Lett. \textbf{111}, 181801 (2013).

\bibitem{LHCb:2013hwj}
R.~Aaij \textit{et al.} [LHCb],
First observation of the decay $B_{c}^{+}\to J/\psi K^+$,
JHEP \textbf{09}, 075 (2013).

\bibitem{LHCb:2014ilr}
R.~Aaij \textit{et al.} [LHCb],
Measurement of the $B_c^+$ meson lifetime using $B_c^+ \to J\!/\!\psi \mu^+ \nu_{\mu} X$ decays,
Eur. Phys. J. C \textbf{74}, 2839 (2014).

\bibitem{LHCb:2014mvo}
R.~Aaij \textit{et al.} [LHCb],
Measurement of $B_c^+$ production in proton-proton collisions at $\sqrt{s}=8$ TeV,
Phys. Rev. Lett. \textbf{114}, 132001 (2015).

\bibitem{LHCb:2017lpu}
R.~Aaij \textit{et al.} [LHCb],
Observation of $B_{c}^{+} \rightarrow D^{0} K^{+}$ decays,
Phys. Rev. Lett. \textbf{118}, 111803 (2017).

\bibitem{CMS:2021vuz}
The CMS Collaboration,
Observation of the ${B_c^+}$ meson in PbPb and pp collisions at $\sqrt{s_{NN}}=5.02~\mathrm{TeV}$,
CMS-PAS-HIN-20-004.

\bibitem{Yang:2011ps}
Z.~Yang, X.~G.~Wu, G.~Chen, Q.~L.~Liao and J.~W.~Zhang,
$B_c$ Meson Production around the $Z^0$ Peak at a High Luminosity $e^+ e^-$ Collider,
Phys. Rev. D \textbf{85}, 094015 (2012).

\bibitem{Yang:2013vba}
Z.~Yang, X.~G.~Wu and X.~Y.~Wang,
BEEC: An event generator for simulating the $B_c$ meson production at an $e^+ e^-$ collider,
Comput. Phys. Commun. \textbf{184}, 2848 (2013).

\bibitem{Chen:2014xka}
G.~Chen, X.~G.~Wu, H.~B.~Fu, H.~Y.~Han and Z.~Sun,
Photoproduction of heavy quarkonium at the ILC,
Phys. Rev. D \textbf{90}, 034004 (2014).

\bibitem{Berezhnoy:2016etd}
A.~V.~Berezhnoy, A.~K.~Likhoded, A.~I.~Onishchenko and S.~V.~Poslavsky,
Next-to-leading order QCD corrections to paired $B_c$ production in $e^+e^-$ annihilation,
Nucl. Phys. B \textbf{915}, 224 (2017).

\bibitem{Zheng:2015ixa}
X.~C.~Zheng, C.~H.~Chang and Z.~Pan,
Production of doubly heavy-flavored hadrons at $e^+e^-$ colliders,
Phys. Rev. D \textbf{93}, 034019 (2016).

\bibitem{Zheng:2017xgj}
X.~C.~Zheng, C.~H.~Chang, T.~F.~Feng and Z.~Pan,
NLO QCD corrections to B$_{c}$(B*$_{c}$) production around the Z pole at an e$^{+}$ e$^{-}$ collider,
Sci. China Phys. Mech. Astron. \textbf{61}, 031012 (2018).

\bibitem{Zheng:2018fqv}
X.~C.~Zheng, C.~H.~Chang and T.~F.~Feng,
A proposal on complementary determination of the effective electro-weak mixing angles via doubly heavy-flavored hadron production at a super Z-factory,
Sci. China Phys. Mech. Astron. \textbf{63}, 281011 (2020).

\bibitem{Chen:2020dtu}
Z.~Q.~Chen, H.~Yang and C.~F.~Qiao,
NLO QCD corrections to $B_c$-pair production in photon-photon collision,
Phys. Rev. D \textbf{102}, 016011 (2020).

\bibitem{CEPCStudyGroup:2018ghi}
J.~B.~Guimar\~aes da Costa \textit{et al.} [CEPC Study Group],
CEPC Conceptual Design Report: Volume 2 - Physics \& Detector,
arXiv:1811.10545 [hep-ex].

\bibitem{An:2018dwb}
F.~An, Y.~Bai, C.~Chen, X.~Chen, Z.~Chen, J.~Guimaraes da Costa, Z.~Cui, Y.~Fang, C.~Fu and J.~Gao, \textit{et al.}
Precision Higgs physics at the CEPC,
Chin. Phys. C \textbf{43}, 043002 (2019).

\bibitem{CMS:2019uhm}
A.~M.~Sirunyan \textit{et al.} [CMS],
Observation of Two Excited B$^+_\mathrm{c}$ States and Measurement of the B$^+_\mathrm{c}$(2S) Mass in pp Collisions at $\sqrt{s} =$ 13 TeV,
Phys. Rev. Lett. \textbf{122}, 132001 (2019).

\bibitem{LHCb:2019bem}
R.~Aaij \textit{et al.} [LHCb],
Observation of an excited $B_c^+$ state,
Phys. Rev. Lett. \textbf{122}, 232001 (2019).

\bibitem{Chang:1992bb}
C.~H.~Chang and Y.~Q.~Chen,
The Production of B(c) or anti-B(c) meson associated with two heavy quark jets in Z0 boson decay,
Phys. Rev. D \textbf{46}, 3845 (1992).

\bibitem{Braaten:1993jn}
E.~Braaten, K.~m.~Cheung and T.~C.~Yuan,
Perturbative QCD fragmentation functions for $B_c$ and $B_{c}$ * production,
Phys. Rev. D \textbf{48}, R5049 (1993).

\bibitem{Ma:1994zt}
J.~P.~Ma,
Calculating fragmentation functions from definitions,
Phys. Lett. B \textbf{332}, 398 (1994).

\bibitem{Chen:1993ii}
Y.~Q.~Chen,
Perturbative QCD predictions for the fragmentation functions of the P wave mesons with two heavy quarks,
Phys. Rev. D \textbf{48}, 5181 (1993).

\bibitem{Yuan:1994hn}
T.~C.~Yuan,
Perturbative QCD fragmentation functions for production of P wave mesons with charm and beauty,
Phys. Rev. D \textbf{50}, 5664 (1994).

\bibitem{Cheung:1995ir}
K.~m.~Cheung and T.~C.~Yuan,
Heavy quark fragmentation functions for $d$ wave quarkonium and charmed beauty mesons,
Phys. Rev. D \textbf{53}, 3591 (1996).

\bibitem{Zheng:2019gnb}
X.~C.~Zheng, C.~H.~Chang, T.~F.~Feng and X.~G.~Wu,
QCD NLO fragmentation functions for c or $\bar{b}$ quark to Bc or Bc* meson and their application,
Phys. Rev. D \textbf{100}, 034004 (2019).

\bibitem{Nason:1993xx}
P.~Nason and B.~R.~Webber,
Scaling violation in e+ e- fragmentation functions: QCD evolution, hadronization and heavy quark mass effects,
Nucl. Phys. B \textbf{421}, 473 (1994).

\bibitem{Dokshitzer:1977sg}
Y.~L.~Dokshitzer,
Calculation of the Structure Functions for Deep Inelastic Scattering and e+ e- Annihilation by Perturbation Theory in Quantum Chromodynamics.,
Sov. Phys. JETP \textbf{46}, 641 (1977).

\bibitem{Gribov:1972ri}
V.~N.~Gribov and L.~N.~Lipatov,
Deep inelastic e p scattering in perturbation theory,
Sov. J. Nucl. Phys. \textbf{15}, 438 (1972).

\bibitem{Altarelli:1977zs}
G.~Altarelli and G.~Parisi,
Asymptotic Freedom in Parton Language,
Nucl. Phys. B \textbf{126}, 298 (1977).

\bibitem{Curci:1980uw}
G.~Curci, W.~Furmanski and R.~Petronzio,
Evolution of Parton Densities Beyond Leading Order: The Nonsinglet Case,
Nucl. Phys. B \textbf{175}, 27-92 (1980).

\bibitem{Furmanski:1980cm}
W.~Furmanski and R.~Petronzio,
Singlet Parton Densities Beyond Leading Order,
Phys. Lett. B \textbf{97}, 437-442 (1980).

\bibitem{Floratos:1978ny}
E.~G.~Floratos, D.~A.~Ross and C.~T.~Sachrajda,
Higher Order Effects in Asymptotically Free Gauge Theories. 2. Flavor Singlet Wilson Operators and Coefficient Functions,
Nucl. Phys. B \textbf{152}, 493-520 (1979).

\bibitem{Gonzalez-Arroyo:1979qht}
A.~Gonzalez-Arroyo and C.~Lopez,
Second Order Contributions to the Structure Functions in Deep Inelastic Scattering. 3. The Singlet Case,
Nucl. Phys. B \textbf{166}, 429-459 (1980).

\bibitem{Floratos:1981hs}
E.~G.~Floratos, C.~Kounnas and R.~Lacaze,
Higher Order QCD Effects in Inclusive Annihilation and Deep Inelastic Scattering,
Nucl. Phys. B \textbf{192}, 417-462 (1981).

\bibitem{Mele:1990cw}
B.~Mele and P.~Nason,
The Fragmentation function for heavy quarks in QCD,
Nucl. Phys. B \textbf{361}, 626 (1991).

\bibitem{Graudenz:1995sk}
D.~Graudenz, M.~Hampel, A.~Vogt and C.~Berger,
The Mellin transform technique for the extraction of the gluon density,
Z. Phys. C \textbf{70}, 77-82 (1996).

\bibitem{Eichten:1994gt}
E.~J.~Eichten and C.~Quigg,
Mesons with beauty and charm: Spectroscopy,
Phys. Rev. D \textbf{49}, 5845 (1994).

\bibitem{ParticleDataGroup:2016lqr}
C.~Patrignani \textit{et al.} [Particle Data Group],
Review of Particle Physics,
Chin. Phys. C \textbf{40}, 100001 (2016).

\end{thebibliography}
\end{document}